\documentclass[sigconf]{acmart}
\usepackage{subcaption}  
\usepackage{setspace}
\usepackage{xcolor, colortbl} 
\usepackage{booktabs}         
\usepackage{tabularx}         
\usepackage{array}            
\usepackage{diagbox}          
\usepackage{threeparttable}   
\usepackage{longtable}        
\usepackage{graphicx}         
\usepackage{caption, subcaption} 

\usepackage{amsmath, amssymb, amsfonts} 
\usepackage[ruled,linesnumbered,vlined,algo2e]{algorithm2e} 

\usepackage{xcolor} 
\usepackage{tcolorbox} 
\usepackage{listings}
\usepackage{balance} 
\usepackage{tikz}
\usepackage{balance}
\usepackage{flushend}      
\usepackage{setspace}
\usepackage{lipsum}        

\lstdefinelanguage{JavaScript}{
  keywords={typeof, new, true, false, catch, function, return, null, catch, switch, var, if, in, while, do, else, case, break, const, prototype},
  keywordstyle=\color{violet}\bfseries,
  ndkeywords={class, export, boolean, throw, implements, import, this},
  ndkeywordstyle=\color{darkgray}\bfseries,
  identifierstyle=\color{black},
  sensitive=false,
  comment=[l]{//},
  morecomment=[s]{/*}{*/},
  commentstyle=\color{gray}\ttfamily,
  stringstyle=\color{blue}\ttfamily,
  morestring=[b]',
  morestring=[b]",
  frame=none,
  numbers=none
}

\hypersetup{
  colorlinks   = true,    
  urlcolor     = blue,    
  linkcolor    = blue,    
  citecolor    = blue      
}

\usepackage{hyperref}
\hypersetup{
  colorlinks   = true,
  urlcolor     = blue,
  linkcolor    = blue,
  citecolor    = blue
}

\usepackage[font=normalsize,skip=2pt]{caption}
\usepackage{xspace} 
\newcommand{\tool}{\textsc{Lap}\xspace}

\newcommand{\tang}[1]{{\color{red} #1}}

\copyrightyear{2025}
\acmYear{2025}
\setcopyright{cc}
\setcctype{by}
\acmConference[FSE Companion '25]{33rd ACM International Conference on the
Foundations of Software Engineering}{June 23--28, 2025}{Trondheim, Norway}
\acmBooktitle{33rd ACM International Conference on the Foundations of
Software Engineering (FSE Companion '25), June 23--28, 2025, Trondheim,
Norway}\acmDOI{10.1145/3696630.3728700}
\acmISBN{979-8-4007-1276-0/2025/06}

\begin{document}

\title{Towards LLM-Based Automatic Playtest}


\author{Yan Zhao}
\email{yzhao19@emich.edu}
\orcid{0009-0003-3908-2664}
\affiliation{%
  \institution{Eastern Michigan University}
  \city{Ann Arbor}
  \state{Michigan}
  \country{USA}
}

\author{Chiawei Tang}
\email{cwtang@vt.edu}
\orcid{}
\affiliation{%
  \institution{Virginia Tech}
  \city{Blacksburg}
  \state{Virginia}
  \country{USA}
}


\begin{abstract}
Playtest is the process in which people play a video game for testing. It is critical for the quality assurance of gaming software. Manual playtest is time-consuming and expensive. However, automating this process is challenging, as playtest typically requires for the domain knowledge and problem-solving skills that most conventional testing tools lack. Recent advancements in artificial intelligence (AI) have opened up new possibilities of applying Large Language Models (LLMs) to playtest. However, significant challenges remain: current LLMs cannot visually perceive game environments; most existing research focuses on text-based games or games with robust API; While many non-text games lack APIs to provide textual descriptions of game states, 
making it almost impossible to na\"ively apply LLMs for playtest.

This paper introduces \tool, our novel approach of \underline{L}LM-based \underline{A}utomatic \underline{P}laytest, which uses ChatGPT to test match-3 games---a category of games where players match three or more identical tiles in a row or column to earn points.  
\tool encompasses three key phrases: processing of game environments, prompting-based action generation, and action execution. Given a match-3 game, \tool takes a snapshot of the game board and converts it to a numeric matrix; it then prompts ChatGPT-O1-mini API to suggest moves based on that matrix; finally, \tool tentatively applies the suggested moves to earn points and trigger changes in the game board. It repeats the above-mentioned three steps iteratively until timeout. 

For evaluation, we conducted a case study by applying \tool to an open-source match-3 game---CasseBonbons, 
and empirically compared \tool with three existing tools. Our results are promising: \tool outperformed existing tools by achieving higher code coverage and triggering more program crashes. Our research will shed light on future research of automatic testing and LLM application.

\end{abstract}



\maketitle

\section{Introduction}

Playtest or game testing is an essential phase in the game development process, ensuring seamless gameplay, validating mechanics, and identifying potential bugs. Despite its significance, playtest remains considerably challenging. Unlike conventional software testing, game testing demands dynamic reasoning, an understanding of spatial relationships, and the ability to simulate diverse player strategies. Traditional testing tools often fall short in mimicking human-like reasoning and adaptability  that game testing necessitates. While reinforcement learning (RL) frameworks offer a promising alternative, their practical adoption faces significant challenges, particularly due to the extensive computational resources and time required for training RL agents. 

The rapid advancement of artificial intelligence (AI) has transformed multiple industries. Among these developments, Large Language Models (LLMs), such as OpenAI's GPT series, have emerged as powerful tools capable of natural language processing, code generation, and even image-based reasoning. This growing intelligence opens new avenues for automating tasks traditionally requiring human cognition, including complex game testing.

In this research, we focus on leveraging Large Language Models (LLMs) for automated play-testing of a typical mobile game genre—Match-3 games\cite{match3games2024}, exemplified by the globally popular Candy Crush Saga\cite{candycrush2023}. These games present a matrix-like board where players swap adjacent tiles to create matches of three or more identical elements, leading to cascading effects and score generation. Despite their apparent simplicity, testing match-3 games involves complex challenges, such as human-like decision-making, spatial reasoning, and adaptive strategies.

However, substantial obstacles persist. Current LLMs struggle with accurate icon positioning in game images due to limited vision capabilities, hindering their spatial reasoning and effectiveness in visually complex tasks like play-testing. Much research has been conducted on text-based games or games with robust API, but these approaches has limited applications. Many non-text games lack APIs that provide text-based game states, making direct interaction challenging, especially in mobile game area. These limitations highlight the need for innovative approaches to bridge this gap and enable LLMs to engage meaningfully in mobile game testing.

In this paper, we present LLM-based Automatic Playtest tool- \tool, the first framework that applies LLMs to play-testing mobile match-3 games. Our approach comprises three key phases:
\begin{itemize}
    \item \textbf{Automated Preprocessing:} Converts visual game board states into structured, interpretable inputs for LLMs.

    \item \textbf{Automatic Prompting Mechanism:} Integrates game rules, examples, and contextual information to guide LLMs in generating valid gameplay actions.

    \item \textbf{Solution Execution:} The outputs of the LLM are translated into executable actions that are evaluated for their validity and impact on gameplay objectives.
\end{itemize}
To evaluate the efficacy of LLM-based LIT, we conduct experiments using CasseBonbons\cite{casseBonbons2020}, an open-source match-3 game inspired by Candy Crush Saga as shown in Figure \ref{fig:gamescreenshot}. Our framework achieves higher code coverage, game scores, game levels and trigger more crushes compared to baseline tools, demonstrating its potential as a robust solution for automated game testing.

This paper is organized as follows: Section II provides a background on match-3 games, the vision capabilities of LLMs and few-shot learning techniques. Section III details the proposed approach, including preprocessing, prompting mechanisms, and solution execution. Section IV presents experimental results and conducting ablation studies. Finally, Section VI discusses future research directions and concludes the study. 
\section{Background}
In this secion, we present the basic concepts of the game, LLM vision capabilities and few-shot learning approach.
\vspace{-4pt}
\subsection{Match-3 Game}
Match-3 puzzle game\cite{match3games2024} is a type of puzzle game where the objective is to match three or more identical items, usually in a row or column, to clear them from the game board. These games typically involve swapping adjacent objects (like candies, gems, or fruits) to create matches on a grid of $m \times n$ tiles.  A well-known example of this genre is Candy Crush Saga developed by King \footnote{\href{https://www.king.com/game/candycrush}{https://www.king.com/game/candycrush}}, which has since been downloaded over 3 billion times as of September 2023, making it one of the most popular mobile games worldwide.\cite{candycrush2023}.

\begin{figure}[t]
    \centering
    \captionsetup{skip=5pt}

    \begin{minipage}[t]{0.2\textwidth}
        \centering
        \includegraphics[width=\textwidth]{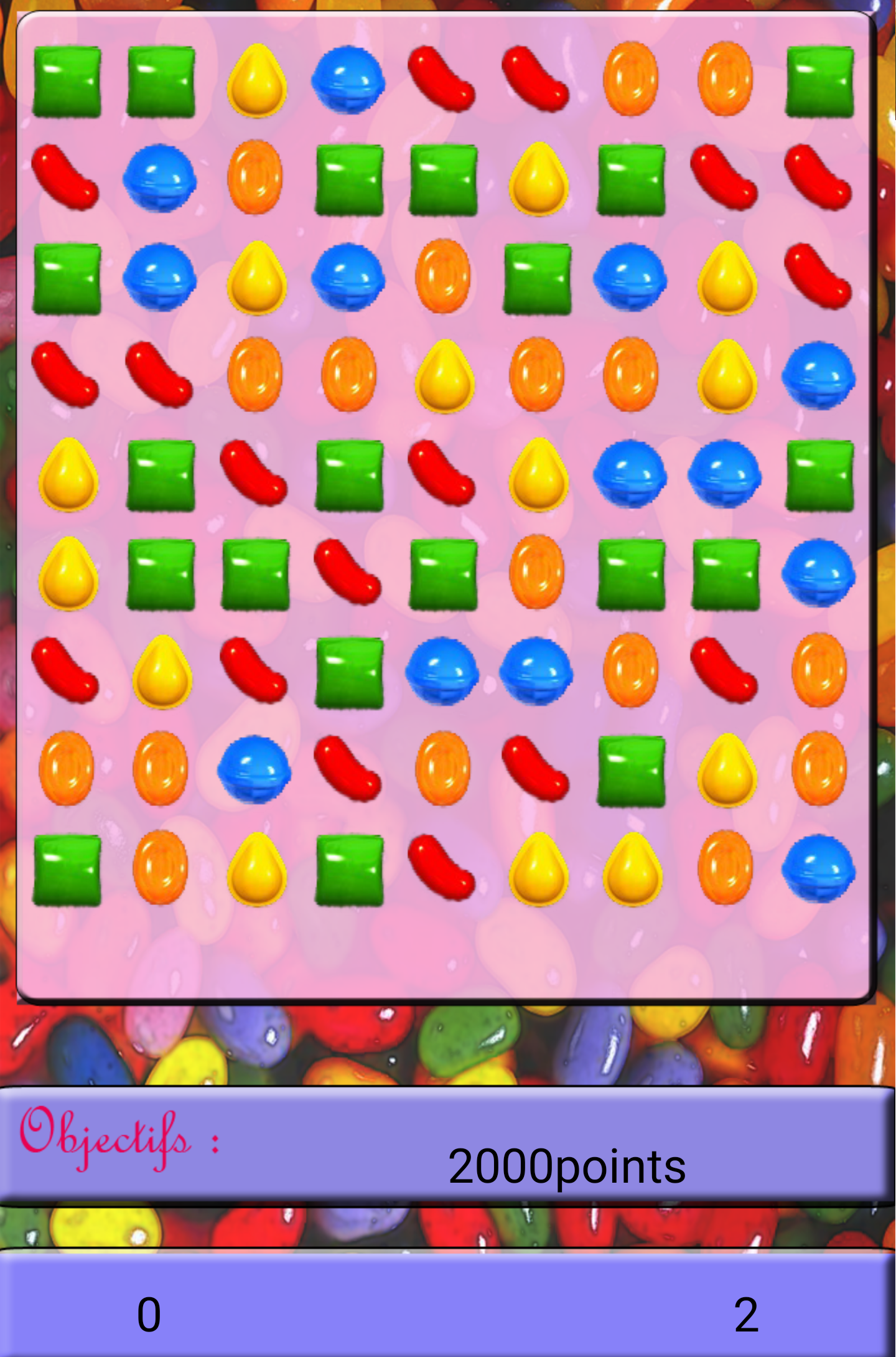}
        \caption{A screenshot of Match-3 Game: CasseBonbons}
        \label{fig:gamescreenshot}
    \end{minipage}
    \hspace{0.05\textwidth}
    \begin{minipage}[t]{0.2\textwidth}
        \centering
        \includegraphics[width=\textwidth]{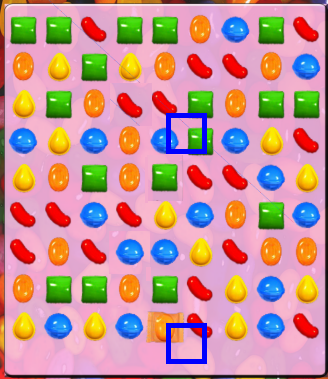}
        \caption{A solution generated by the LLM using vision capabilities}
        \label{fig:marked_candy}
    \end{minipage}
\end{figure}

In this study, we chose a match-3 game as the basis for our experiments due to its widespread popularity and well-documented gameplay mechanics. And we utilize CasseBonbons~\cite{casseBonbons2020}, an open-source implementation inspired by Candy Crush Saga as shown in Figure \ref{fig:gamescreenshot}, which allowed us to effectively collect test results and evaluate the efficiency and validity of our testing methodologies.

\vspace{-4pt}
\subsection{LLM VISION}


Modern LLMs, such as OpenAI's GPT-4\cite{openai2023gpt4}, incorporate multimodal capabilities that enable visual content processing and understanding. While these models excel at general image comprehension and object identification, they exhibit notable limitations in spatial reasoning tasks. According to OpenAI's technical documentation\footnote{\href{https://platform.openai.com/docs/guides/vision?lang=node}{OpenAI Vision Documentation}\label{openai-vision-doc}}, the models perform well when answering broad questions about image content and understanding object relationships. However, they demonstrate reduced accuracy when processing queries that require precise spatial localization. For instance, while the model can reliably identify object attributes (e.g., the color of a vehicle) or suggest contextual actions (e.g., meal recommendations based on refrigerator contents), it struggles with spatial queries such as determining the specific location of furniture within a room layout.

To evaluate these capabilities, we conducted a preliminary test by providing multiple screenshots of a Match-3 game board to OpenAI's GPT-4. We provide the model with rules and ask it to generate a solution for playtesting. However, the model did not produce accurate responses as expected.  This limitation aligns with the observations detailed in the OpenAI's technical documentation\tang{\footref{openai-vision-doc}}.


Figure~\ref{fig:marked_candy} illustrates a representative case where GPT-4 was tasked with identifying optimal candy-swapping moves given a game board configuration and rule set. The model exhibited consistent difficulty in both precise candy localization and strategic move generation, underscoring fundamental limitations in its spatial reasoning capabilities. These constraints substantially impact the model's ability to autonomously analyze complex game states and generate optimal gameplay strategies based solely on visual input, suggesting that current vision-language models may require additional architectural innovations to handle spatially-dependent decision-making tasks effectively. 

\vspace{-4pt}
\subsection{Few-Shot Learning Approach}

To address these limitations, we leverage few-shot learning\cite{vinyals2016matching}, a machine learning paradigm that enables models to generalize from a small set of examples. 
Our methodology applies this principle by providing the LLM with human-generated examples of optimal candy-crushing moves. These carefully curated examples serve to demonstrate game mechanics and valid move patterns, enabling the model to generate contextually appropriate actions based on the current game state.
More details about the few-shot learning 
application in this work will be discussed in the \textit{Approach} chapter.

\lstset{
numbers=left,
basicstyle=\scriptsize,
numberstyle=\scriptsize,
breaklines=true,
numbersep=1.5pt,
frame=tb}

\section{Approach}
\begin{figure*}[ht]
    \centering
   {\includegraphics[width=0.6\paperwidth]{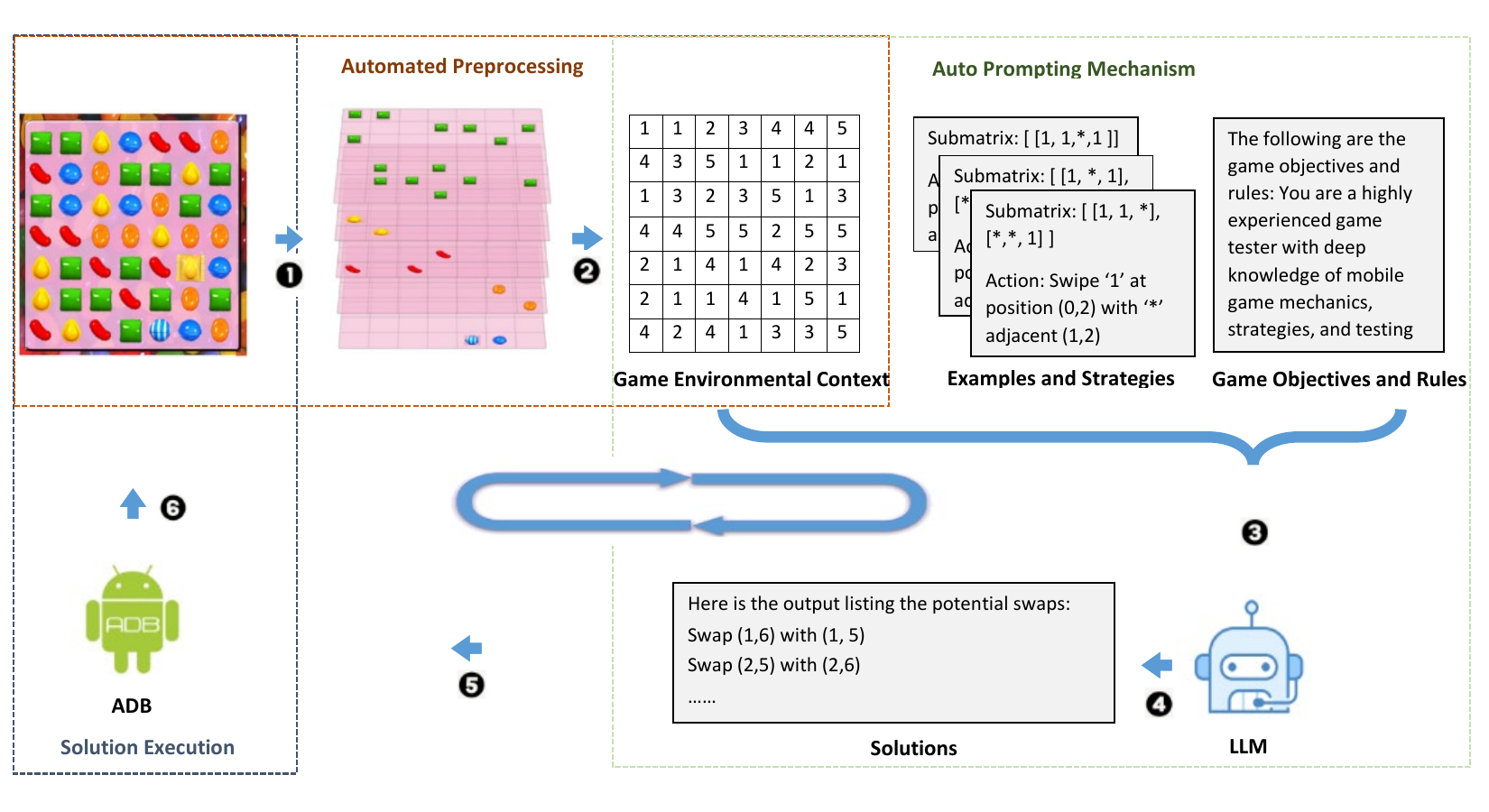}}
    \caption{Framework of LAP}
    \label{fig:framework}
\end{figure*}

In this section, we present \tool to address these limitations. Specifically, we outline a method to leverage LLMs for playing mobile games by employing a combination of automated preprocessing, autoprompting mechanism, and soluciton execution. 
\vspace{-4pt}
\subsection{Automated Preprocessing}

\tool begins by capturing a screenshot of the game board using the command-line tool Android Debug Bridge (ADB)~\cite{GoogleADB} from the Android device. For each screenshot, \tool sequentially recognizes the specified candy icons using OpenCV (Open Source Computer Vision Library) \cite{opencv2020}, which outputs their corresponding coordinates on the board.  Based on the positions and counts of the detected candy icons, the match-3 game board is represented as an $m \times n$ matrix, where each cell corresponds to a position occupied by a candy of a specific color. As illustrated in Step~1 of Figure~\ref{fig:framework}, the raw game scene is decomposed into multiple feature matrix layers, with each layer corresponding to a specific candy color. 

For each game-playing iteration, every candy's color is represented by a unique number. The corresponding feature layer for the candy color is converted into a numerical matrix. After processing all layers, as shown in Step 2 of Figure ~\ref{fig:framework}: The game board will be transformed into a 2D matrix, in which each cell is assigned the number corresponding to the candy's color.

The action of swapping two adjacent candies on the game board is abstracted as the exchange of values between two adjacent cells in the 2D matrix. This abstraction simplifies the representation of moves, making it more interpretable for downstream processing. 

\begin{figure}[ht]
    \centering
    \includegraphics[width=0.25\textwidth]{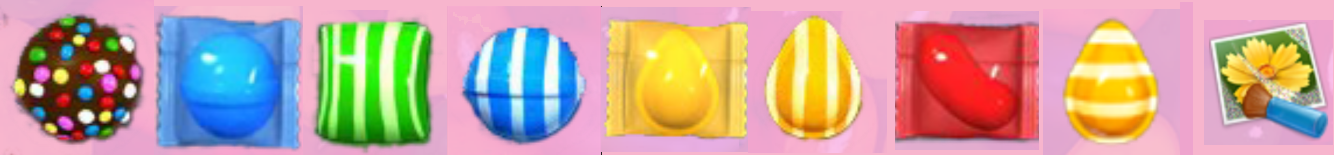}
    \caption{Example of Special Candies and Blocker Candies in Match-3 Game CasseBonbons}
    \label{fig:special_candy}
\end{figure}

In addition to regular candies, Match-3 games usually incorporate special candies and blocker candies (see \ref{fig:special_candy}) .  
Special candies are generated by matching at least four candies of the same color. And each special candy has a distinct effect that can be activated upon being smashed. The leftmost special candy is the color bomb (represented by number 0), which can be swapped with any adjacent candy to eliminate all candies of the same color. The rightmost special candy is the blocker candy in Casse-Bonbons (represented by -1), which cannot be swapped. 
\vspace{4pt}

\vspace{-8pt}
\subsection{Automatic Prompting Mechanism}

The prompt framework provided to the LLM is structured as follows. For the game being tested, we replace the placeholders accordingly:

"You are given a game to play. The game's environmental context is: \{ Game Environmental Context \}.
Here are examples and strategies you can refer to while playing the game: \{ Examples and Strategies \}.
The following are the game objectives and rules: \{ Game Objectives and Rules \}."

As shown in Step~3 of Figure~\ref{fig:framework}, we will demonstrate how to define the placeholders using the game CasseBonbons.

\begin{enumerate}
    \item \textbf{Game Environmental Context} This part of the prompt contains the game board context, preprocessed using the framework described in the previous section. Our tool continuously updates the LLM with current game state.
    \begin{lstlisting}[language=python, caption=Game Environmental Context Prompt,label=lst:rules]
[{'role': 'user', 'content': "<task> The game's environmental context is a 2D matrix as follows: 
    [[3, 1, 3, 5, 1, 1, 3, 3, 1],
     [1, 3, 4, 3, 3, 1, 5, 3, 5],
      ...
  </task>"}]
    \end{lstlisting}
    \item \textbf{Examples and Strategies} 
The middle part of the prompt provides specific examples to the LLM as part of a few-shot learning approach to module new solutions. In this paper, we supplied 15 sets of sub-matrices and corresponding actions. We define an action as swapping two adjacent candy of the game board. Each example demonstrates an action performed on a submatrix that results in the crushing of candies. 

For instance, one set of submatrix and action provided in the example prompt is as follows (All matrices are 0-indexed.). The symbol * serve as a wild card that can represent any types of candy, and does not need to remain the same in one sub matrix. The symbol \texttt{1} must consistently represent the same letter (regardless of case)  corresponding to one specific candy color throughout the submatrix.


\begin{itemize}
    \item \textbf{Submatrix:}
    \[
    \begin{bmatrix}
    * & \tikz[baseline=(X.base)] \node[draw, circle, inner sep=0.3pt] (X)  {\makebox[1em]{$1$}}; & * \\
    1 & \tikz[baseline=(X.base)] \node[draw, circle, inner sep=0.3pt] (X) {\makebox[1em]{$*$}}; & 1
    \end{bmatrix}
    \]
    \item \textbf{Action:} Swipe the value \texttt{1} at position $(0, 1)$ with the adjacent \texttt{*} at position $(1,1)$.
\end{itemize}

After the swipe action, the submatrix is transformed into:

\[
\begin{bmatrix}
* & * & * \\
1 & 1 & 1
\end{bmatrix}
\]

In the transformed submatrix, the second row will crush as it contains three candies of the same color in a row.

These structured example sets will be fed into the LLM to illustrate the rules of candy crushing.

\begin{lstlisting}[language=python, caption=Game Example Prompt,label=lst:examaples]
[{'role': 'user', 'content': "The <example> tag provides examples of the game playtest, showing a specific submatrix pattern of candies and an exact swipe action.Please reference those submatrix pattern in the game test. And make sure the entire submatrix must be matched as it appears in the examples in the game matrix. Moreover, because actions in examples use coordinates based on the submatrix..."}]
[{'role': 'user', 'content': "
      <example>Submatrix:[['1', '1', '*', '1']], Action: swipe '1' at position (0,3) with adjacent '*' at position (0,2) )</example>
      <example>Submatrix:[['*', '1', '1'], ['1', '*', '*']], Action: swipe '1' at position (1,0) with adjacent '*' at position (0,0))</example>
      ...}]"
\end{lstlisting}

\item \textbf{Game Objectives and Rules:} The final part of the prompt poses rule queries.

\begin{lstlisting}[language=python, caption=Rule Prompt,label=lst:rules]
[{'role': 'user', 'content': "You are a highly experienced game tester with deep knowledge of mobile game mechanics, strategies, and testing methodologies.Your role is to follow game rules and meet specific testing requirements to help assess the game's functionality and enjoyment..."}] 
[{'role': 'user', 'content': 'The <rule> tag provides all game rules. These rules define the game mechanics, objectives, constraints, and any special actions allowed in the game.
    <rule>You are testing a game based on the mechanics of Match-3 games. Players swap adjacent candies to form horizontal or vertical lines of three or more candies of the same color. A match of three or more clears those candies from the board.</rule>
    <rule> The game board has been preprocessed and represented as a 2D matrices, with numbers representing candies of specific colors and types.</rule>
    <rule>Color bombs(represented by 0) are special candies and should be swapped with any adjacent candy to amximize the effect. These should be prioritiezed for swaps where possible</rule>
...'}] 
\end{lstlisting}
\vspace{-.5em}
 
\end{enumerate}

Ultimately, \tool will prompt the LLM to recommend action sets based on the provided game matrix, rules, and examples.
\vspace{-4pt}
\subsection{Solution Execution}

 We process the set of actions provided by the LLM and convert them into executable commands to control the game. Specifically, we use ADB~\cite{GoogleADB} to control the Android device. ADB provides access to a Unix shell, allowing us to execute various commands on an Android device. First, we transform the matrix coordinates generated by the LLM into game screen coordinates that ADB can interpret to control the phone screen. For example, the action swipe(1,6) with (1,5) is converted into swipe(123px,1098px) with (123px,1370px). Then, we issue the corresponding swipe command via ADB to interact with the Android phone:
\begin{lstlisting}
adb shell input swipe 123 1098 123 1370
\end{lstlisting}

\section{Experiments}

In this section, we first describe the experimental settings and then explain the conducted experiments and analyze the
results.

\vspace{-4pt}
\subsection{Experimental Settings}

We leverage OpenAI's GPT-O1-mini APIs for playtesting. The API was called using Python 3.10 in a Linux-based environment with the openai Python package. We configured the model with a temperature of 0.5 to prioritize response consistency, and limited the output using max\_tokens = 500 and num\_samples = 1 to control response length and determinism. Our simulation environment is built on top of Android Emulator provided by Android Studio and leverages ADB toolset for mobile game control. 

\vspace{-4pt}
\subsection{Baseline}
Due to the absence of LLM-based testing tools for playtesting mobile games, we selected a number of rule-based, randomness-based testing and Reinforcement Learning-based tools as baseline approaches for comparison.

\begin{itemize}
    \item \textbf{Monkey Tool}\cite{monkey2020} is a randomness-based testing framework commonly used in mobile game and application testing. It generates random input events, such as touches, gestures, and presses, and sends them to the application under test. 
    
    \item \textbf{LIT Tool}\cite{9825886} is a rule-based Lightweight Approach of Human-Like Playtest, which could generalize playtest tactics from manual testing, and to adopt the tactics for automatic testing  Android Apps. Given the match-3 game G, the first author manually played G for eight minutes in LIT's demo mode, and then switched to LIT's test mode to automatically play G. We use the tool as baseline in our experiment, but test the match-3 game for 150 iterations instead of one hour in the original paper.

    \item  \textbf{RLT}\cite{9825886} is a vanilla Reinforcement Learning-based tool built in LIT paper and we refer to it as RLT. We trained RLT for eight minutes (i.e., the same length with the LIT demo time) aligning its training duration with the LIT demo time to ensure consistency across all tools.
\end{itemize}
\vspace{-4pt}
\subsection{Metric}
We evaluated the effectiveness of various testing tools by measuring code coverage and bug detection during execution. The general assumption is that a testing tool's effectiveness increases as it executes more code. To quantify coverage, we used line coverage obtained by JaCoCo \cite{eclemma2023}.

Since our experiment target is mobile games, we introduced two additional metrics: \textbf{Game Score} and \textbf{Game Level}. The \textbf{Game Score} represents the points earned by a testing tool after it interacts with the game for a set period. And the \textbf{Game Level} indicates the number of levels the tool achieves before the testing time expires; a higher score and level suggests a tool unlock more game scenarios and therefore, more thorough testing.

\vspace{-4pt}
\subsection{Experiment Results}
We systematically evaluate \tool and baselines on their play-testing performance across 150 iterations.

\begin{figure}[ht]
    \centering
    \includegraphics[width=0.42\textwidth]{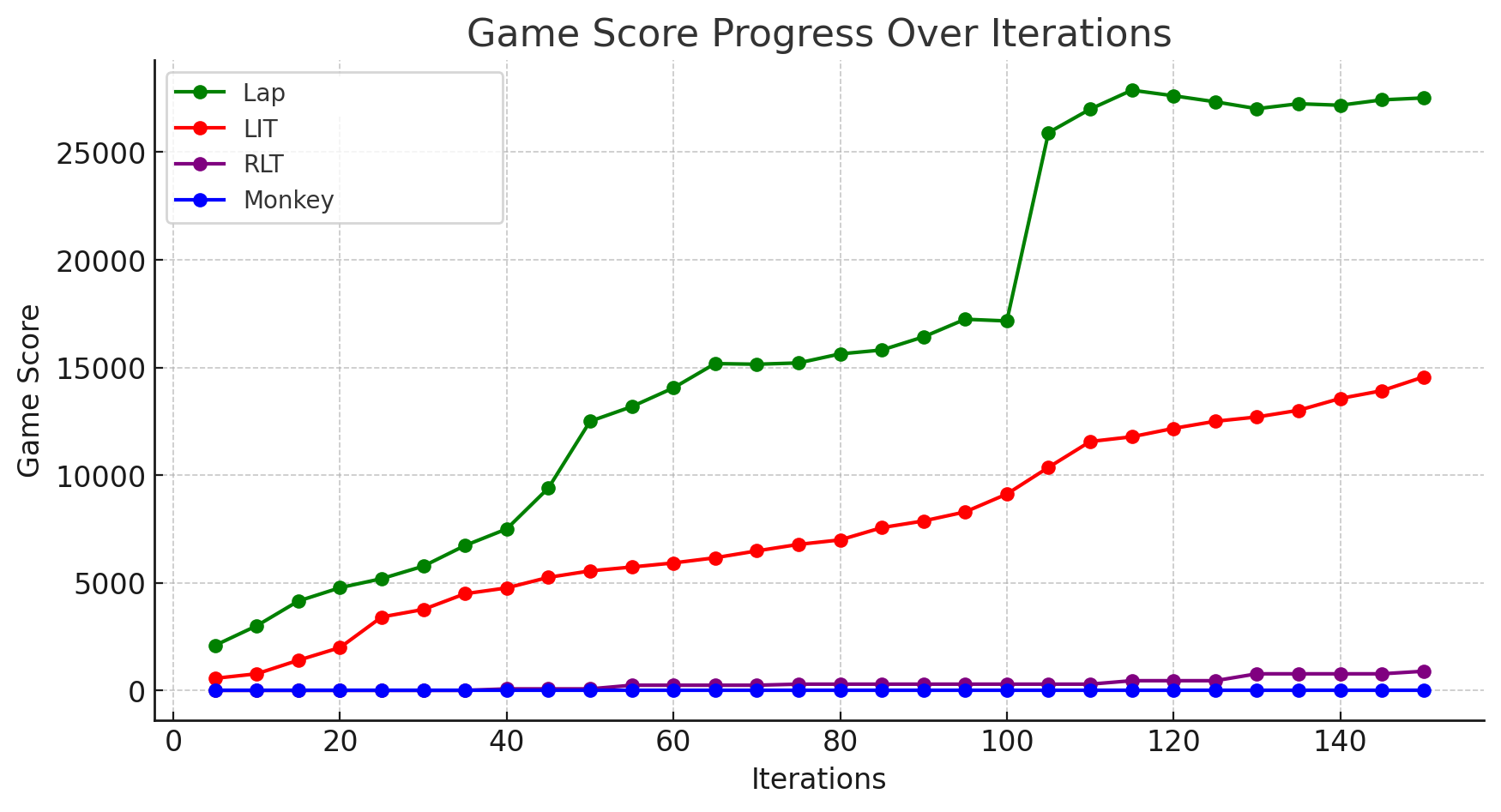}
    \caption{Game Scores Across Iteration Rounds}
    \label{fig:all_scores}
\end{figure}

\begin{figure}[ht]
    \centering
    \includegraphics[width=0.42\textwidth]{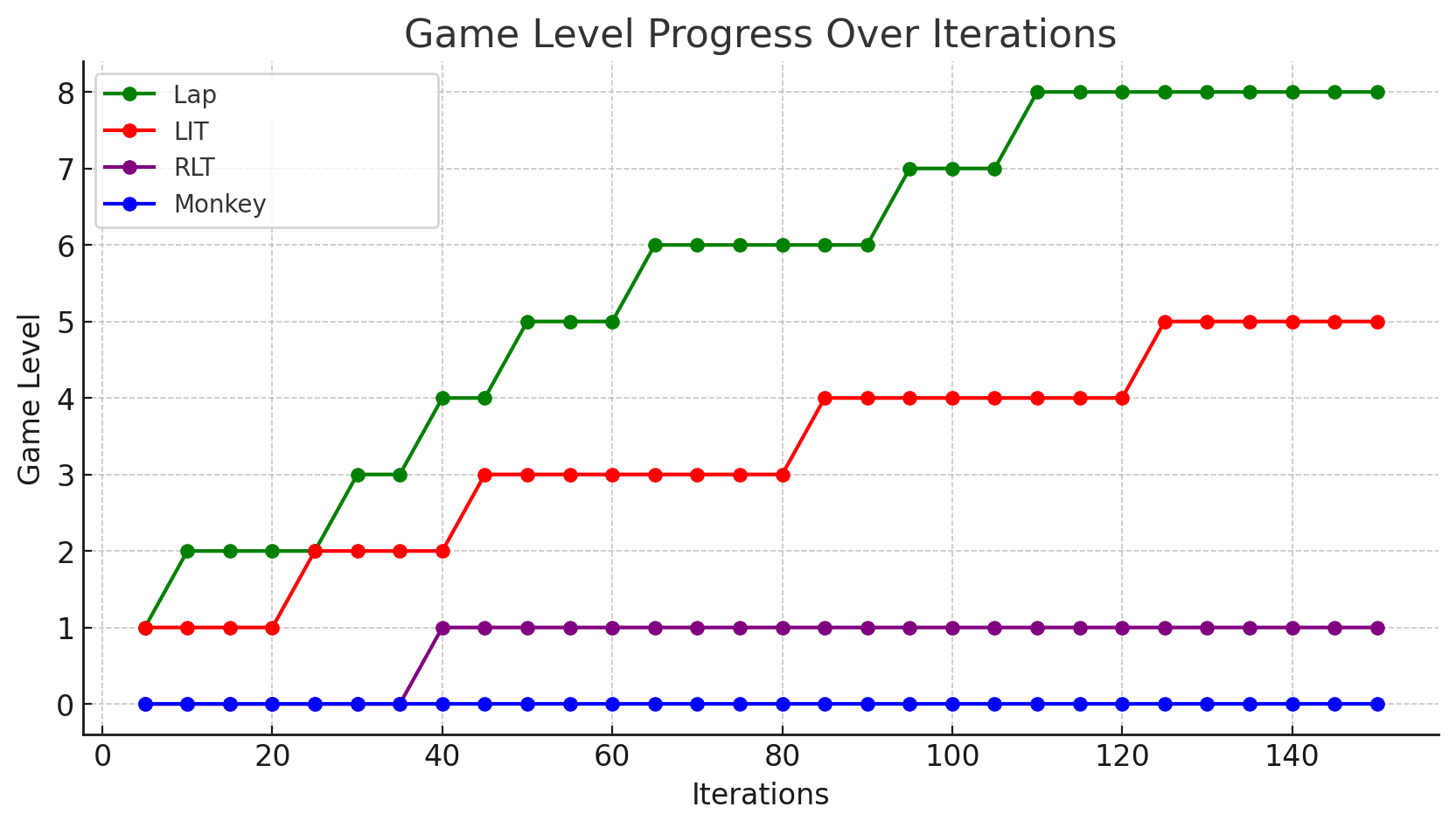}
    \caption{Game Levels Across Iteration Rounds}
    \label{fig:all_levels}
\end{figure}

\vspace{2pt}

\textbf{Extensive Testing Scope}:
The game score results are summarized in figure \ref{fig:all_scores}, which shows the scores achieved by each tool over iterations. The game level progress are shown in \ref{fig:all_levels}.  

In our experiments, \tool achieves the highest score of 27,520 and highest level of 8,  which demonstrates steady growth across iterations. The "LIT" tool shows consistent improvement over iterations but at a slower pace, reaching a peak score of 14,560 and level of 5. "RLT" starts at 0 and remains low for the majority of iterations. As explained in LIT paper\cite{9825886}, "As we trained RLT for only eight minutes (i.e., the same length with the LIT demo time), it is possible that RLT was not trained sufficiently and it worked less effectively". The "Monkey" category remains constant at 0 throughout all iterations because it does not know how to enter the game, and spent lots of time clicking random pixels
on the display before accidentally hitting the start button.

This result highlights \tool, which use the LLM as the cognitive engine, can understand game mechanics, generate targeted game testing scenarios, and conduct effective game testing.

\vspace{4pt}
\textbf{Enhanced Code Exploration}:
\begin{figure}[ht]
    \centering
    \includegraphics[width=0.42\textwidth]{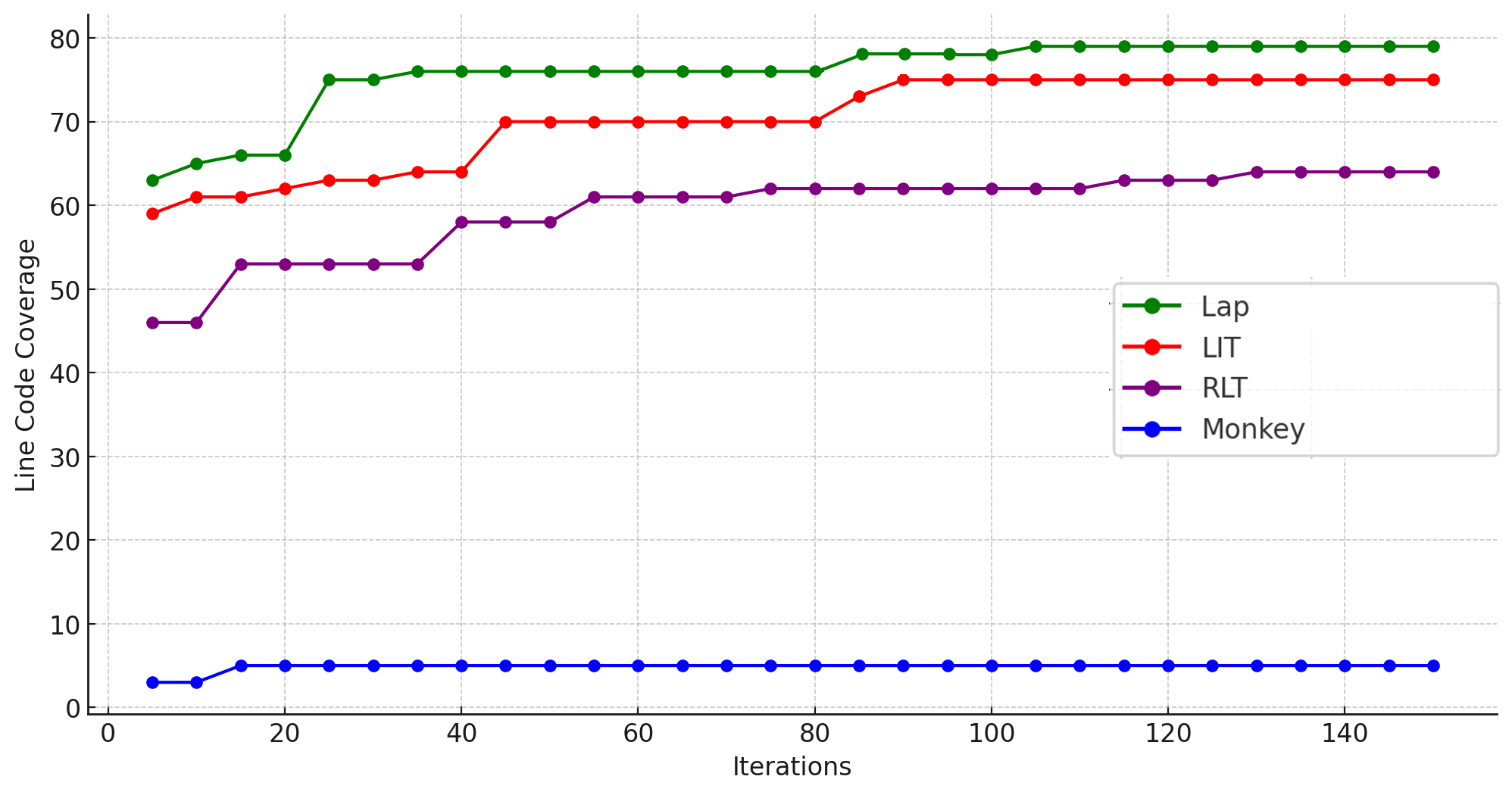}
    \caption{Line Code Coverage Across Iteration Rounds}
    \label{fig:all_coverage}
\end{figure}
The code coverage results are summarized in Figure \ref{fig:all_coverage}, which shows the line coverage percentage achieved by each tool. In our experiments, \tool shows continuous improvement in coverage initially and achieved the highest values 79\%. The "LIT" series shows consistent growth over iterations but at a slower pace. it stable at 75\% until the final iteration. "RLT" starts at 46\% and shows steady progress but achieves a lower final coverage compared to LIT and LLM. The "Monkey" category maintains an almost constant coverage of 3-5\% across all iterations, exhibiting a flatline with negligible improvement.

This comprehensive exploration ensures that most of the codebase is exercised during testing, minimizing blind spots where potential bugs might hide.

\vspace{2pt}
\textbf{Improved Vulnerability Discovery}:
During testing of Cassebonbons with \tool, the game crashed as shown in Table~\ref{tab:crush}, indicating its ability to uncover previously hidden vulnerabilities.

These crashes suggest that the LLM tool explores deep and unexpected paths in the application. By uncovering these crashes, the LLM tool provides developers with actionable insights to improve the robustness and stability of their applications.
\vspace{4pt}
\begin{table}[h!]
\centering
\footnotesize
\caption{ Crash Triggering Results}
\label{tab:crush}
\begin{tabular}{lc} 
\toprule
\textbf{}      & \textbf{Number of crush triggering} \\ 
\midrule
LAP            &5                          \\
LIT            &1                          \\
Monkey         &0                           \\
RLT            &0                          \\

\bottomrule
\end{tabular}
\end{table}

\vspace{-5pt}
\subsection{Ablation Studies of \tool}

\begin{figure}[ht]
    \centering
    \includegraphics[width=0.42\textwidth]{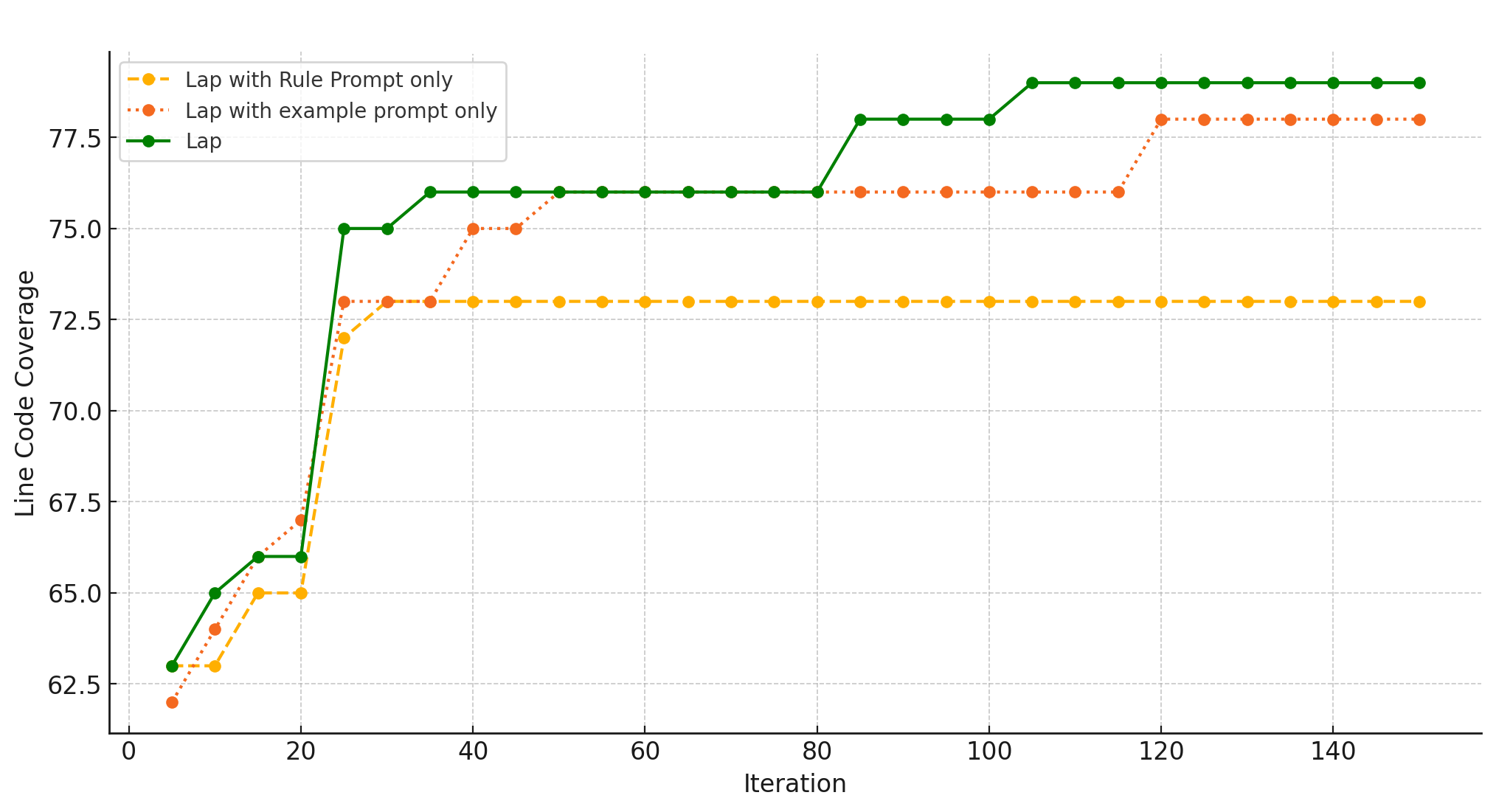}
    \caption{Line Code Coverage Across Iteration Rounds}
    \label{fig:coverage}
\end{figure}

We ablate 3 input prompt choices, \tool with rule prompt only, \tool with example prompt only and \tool with both rule prompt and examples as Few-shot Learning prompt. We study their impact on playtesting performances within 150 prompting iterations. Results are show in \ref{fig:scores} and \ref{fig:coverage}. X-axis denotes the number of prompting iterations. We highlight the key findings below:

\begin{figure}[ht]
    \centering
    \includegraphics[width=0.42\textwidth]{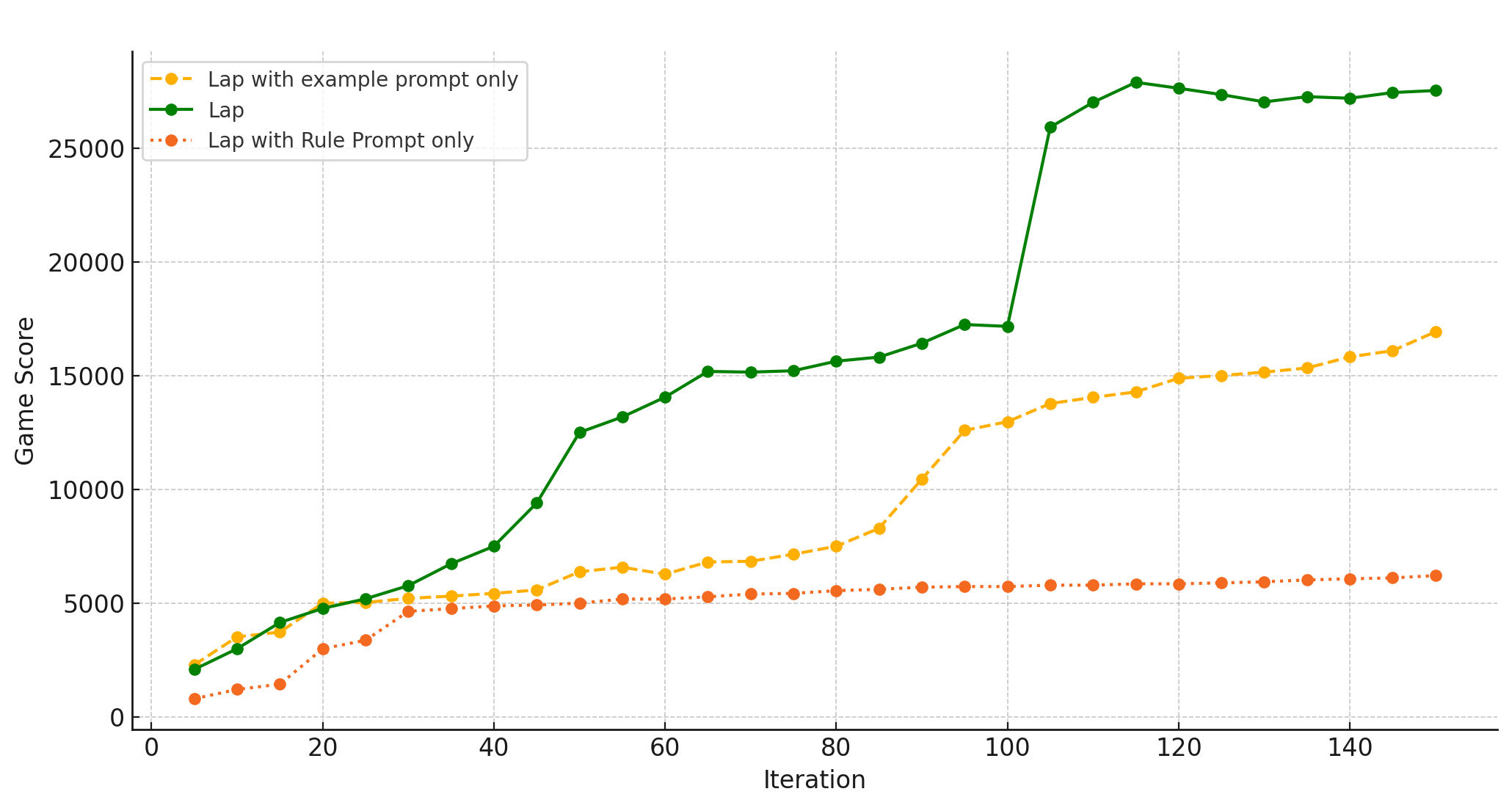}
    \caption{Game Scores Across Iteration Rounds}
    \label{fig:scores}
\end{figure}

\begin{itemize}
    \item \tool with rule prompt alone enables game play-testing but lacks efficiency. The overall game progression is relatively slow. 
    \item \tool with Few-shot Learning alone provides a notable improvement, demonstrating the importance of examples in enhancing play-testing performance.
    \item The combination of Rule Prompt and Few-shot Learning significantly proves to be the most effective for both short-term progress and long-term optimization, highlighting the synergy between rule prompts and few-shot learning, where the structured guidance from rules complements the adaptability of few-shot learning.
\end{itemize}
\section{Related Work}
\subsection{Automated Testing for Android Apps}
Various tools were proposed to automate testing for Android apps. Specifically, random-based approaches (e.g., DynoDroid \cite{10.1145/2491411.2491450} and Monkey \cite{monkey2020}) tests app by generating random UI events and system events. Given an app to test, model-based approaches (e.g. GUIRipper \cite{6494930}, Time travel \cite{10.1145/3377811.3380402}, EvoDrodi \cite{10.1145/2635868.2635896} and PUMA \cite{10.1145/2594368.2594390}) use static or dynamic program analysis to build a model for the app as a finite-state machine (FSM). Record-and-replay tools (e.g., RERAN \cite{6606553} and  Mobi-  Play \cite{10.1145/2884781.2884854}) record inputs and program execution when users manually test apps, and then replay the recorded data to automatically repeat the testing scripts. 

However, these approaches are not well-suited for mobile game testing due to the complexity of game interactions, real-time execution constraints, high state variability, performance dependencies, and the lack of standardized UI components. These unique challenges require specialized testing tools tailored for game apps.
\vspace{-4pt}
\subsection{Automated Game Testing}
Several approaches were introduced to automate game testing. Specifically, online testing (e.g., TorX \cite{16cb22c0962c42349e8846d75f5186c6} and Spec Explorer \cite{10.1145/1081706.1081751}) relies on model-based testing to detect discrepancies between the implementation under test (IUT) and a formal specification. However, these methods require informal specification languages of expertise, while \tool does not.

 Deeplearning-based approaches use large-scale player data to train models that predict human-like actions in gameplay. Specifically, Bergdahl et al. (2021)\cite{9231552} proposed a framework that utilizes DRL to explore and exploit game mechanics based on user-defined reward signals.  Reinforcement learning-based approaches utilize reward systems to teach agents through trial and error, allowing them to learn optimal strategies by interacting with the game environment and receiving feedback on their actions. Wuji \cite{8952543} integrates evolutionary algorithms, DRL, and multi-objective optimizations for automated game testing. However, DL-based \cite{pham2020atgw},\cite{9231552},\cite{10.1109/CIG.2018.8490442}, \cite{10.1145/3472672.3473952} or RL-based tools \cite{10069395} \cite{8952543} cannot be directly applied to new games without significant retraining, requiring extensive data and computational resources. In contrast, Lap leverages LLM as its core engine, eliminating the need for vast datasets and intensive training, making it more adaptable and efficient for game testing. 
\vspace{-4pt}
\subsection{LLM-Driven Game Playing}
Large language models (LLMs)-driven automated game playing has emerged as a transformative approach. We identify
two general classes of games to which LLM players are
well suited: 

(a)
Games where the main input and output modalities are natural
language. For instance, HuMAL\cite{10.1609/aaai.v38i19.30155} trains agents to  improved decision-making and planning for text-based game. CALM \cite{yao-etal-2020-keep} is a GPT-2 system finetuned on a dataset of human
gameplay transcripts collected from a variety of text adventure
games. The GALAD system
\cite{ammanabrolu-etal-2022-aligning} uses a pre-trained LLM to guide an agent towards
morally acceptable actions in text games from the Jericho
suite \cite{hausknecht2020interactive}. 

(b) Games for which external programs
can control player actions via an API.  VOYAGER \cite{wang2023voyager} leverages the code generation abilities of GPT-4
to play Minecraft (Mojang Studios, 2011) by interacting with
the popular Mineflayer API. MineDojo \cite{fan2022minedojo} and VPT \cite{10.5555/3600270.3602059} utilize YouTube videos for large-scale pre-training to play games through API.

 However, many non–text-based games, including mobile games, do not offer robust API support for retrieving game states and controlling gameplay. This limitation significantly restricts the direct application of existing approaches, motivating alternative solutions such as the one proposed in \tool{}.

\section{Limitation and Future Research Directions}

\textbf{Hallucinations}:Large Language Models (LLMs) can sometimes generate invalid solutions, failing to make further progress. To mitigate the impact of hallucinations, we provide LLMs with as much game information as possible, enabling them to return the most extensive set of potential results. 

\textbf{Time Sensitivity}:Some games require players to provide a solution within a specific time limit, such as a car racing game. Since LLMs require time to think and generate responses, they may not be well-suited for tasks with strict time constraints.

\textbf{Scalability}: Although we implemented mechanisms to transform the match-3 game environment and state into a format understandable by LLMs and enable LLMs to play-test the match-3 game effectively. However, whether similar transformations and effective testing can be achieved for other mobile games remains a topic for further research.

\section{Conclusion}
In this study, we introduced \tool, a novel framework to automate play testing in match-3 mobile games using LLMs. Using automated pre-processing, a structured prompting mechanism, and solution execution, our framework enables LLMs to interpret game states, generate valid actions, and achieve predefined gameplay objectives.

Experimental results demonstrated that \tool outperforms conventional baseline tools in terms of code coverage, game score, and vulnerability discovery. Furthermore, our ablation studies highlighted the synergy between rule prompts and few-shot learning, showcasing how combining structured guidance with adaptive examples enhances performance.

In conclusion, \tool  demonstrates the potential of AI in decision-making and
actions within complex game environments, bridges the gap between advanced artificial intelligence and automated play-testing, provide a robust and innovative solution for mobile game testing. This work paves the way for future research into LLM-driven frameworks for other game genres and complex testing environments, contributing to the broader adoption of AI in game testing. 

\begin{spacing}{1.1}
\raggedbottom
\bibliographystyle{plainnat}

\bibliography{main}
\end{spacing}
\end{document}